\documentclass[twocolumn,twoside,slac]{revtex4}
\usepackage{graphicx}
\usepackage{fancyhdr}
\begin{document}

\title{A ROOT-based Client-Server Event Display for the ZEUS Experiment}

%

%
\author{O. Kind, J. Rautenberg}
\affiliation{University of Bonn,Physikalisches Institut,D-53115 Bonn, 
Germany}
\author{C. Genta} 
\affiliation{INFN and Dipartimento di Fisica dell'Universita` di
  Firenze, I-50019 Sesto Fiorentino (Florence), Italy}
\author{S. Hanlon} 
\affiliation{Dept of Physics and
  Astronomy,University of Glasgow, Glasgow, G12 8QQ, United Kingdom}
\author{U. Fricke, O. Gutsche, R. Kaczorowski, R. Mankel, K. Wrona}
\affiliation{DESY Hamburg, D-22603 Hamburg, Germany}
\author{E. Heaphy}
\affiliation{Pennsylvania State University, Dept. of Physics,
University Park, PA 16802 , USA}
\begin{abstract}
  A new event visualization tool for the ZEUS experiment is nearing
  completion, which will provide the functionality required by the new
  detector components implemented during the recently achieved HERA
  luminosity upgrade. The new design is centered around a
  client-server concept, which allows to obtain random access to the
  ZEUS central event store as well as to events taken online via the
  HTTP protocol, even from remote locations. The client is a
  lightweight C++ application, and the ROOT system is used as
  underlying toolkit. Particular attention has been given to a smooth
  integration of 3-dimensional and layered 2-dimensional
  visualizations. The functionality of server and client application
  with its graphical user interface are presented.
\end{abstract}

\maketitle

\thispagestyle{fancy}


\section{Introduction}
Event displays are indispensable tools for analysis in high energy
physics experiments. They help to understand the physics of a recorded
interaction, to diagnose the apparatus, to make the detector geometry
imaginable and to illustrate the whole matter to general audiences.
For these rather heterogeneous requirements, the application must be
versatile and offer both 2-dimensional (2D) and 3-dimensional (3D)
representations. On the other hand, in today's large collaborations'
distributed environments, it should be possible to obtain event
displays regardless of the question whether the workstation is on site
or somewhere else in the world.  This is in general a challenge, since
displaying events requires access to the experiment's event store and
geometry constants.

In the ZEUS experiment, new requirements have emerged with the
installation of new detector components, in particular the silicon
micro-vertex detector (MVD) and the forward straw tube tracker (STT),
which require more elaborate 3-dimensional visualizations than the
previous tool provided.

In the following, we describe the {\it ZeVis} program suite (ZeVis
standing for {\bf ZE}US {\bf Vis}ualization), which is a distributed
application designed to meet the above challenges.

\section{Design Criteria}
The design of the new event display is governed by two key criteria
which are detailed in the following.

\subsection{Client Server Capability}
The system should consist of a highly portable light-weight client,
capable of running on many different platforms with the host residing
on-site or somewhere in the internet, and a server application which
lives on a stationary host in direct contact with the event store. The
term ``light-weight'' implies that the client should be independent of
the huge legacy libraries making up the main reconstruction and
analysis code. The server application, on the other hand, should
incorporate standard analysis tools. The connection protocol between
client and server should be standardized and likely not to be stopped
by firewalls.

\subsection{Integration of 2D/3D Views}
\begin{figure}
\includegraphics[width=80mm]{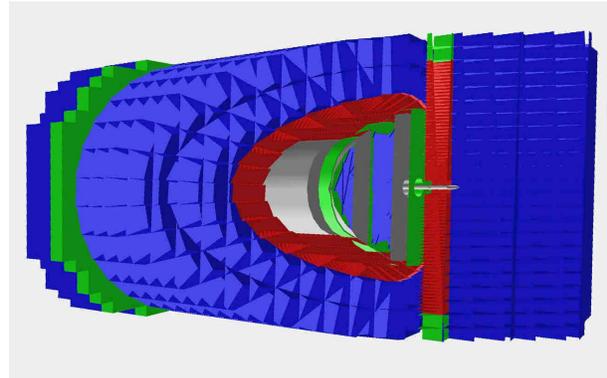}
\caption{3-dimensional view of detector geometry with hidden lines removal.}
\label{fig:detector-3D}
\end{figure}
Perspective 3-dimensional views with hidden lines and hidden surface
removal are very useful for understanding detector geometry, and
provide attractive pictures for public relation purposes
(fig.~\ref{fig:detector-3D}).  Analyzing the event itself is often
less successful in this mode, since the geometry tends to obscure the
tracks and hits.

\begin{figure}
\includegraphics[width=80mm]{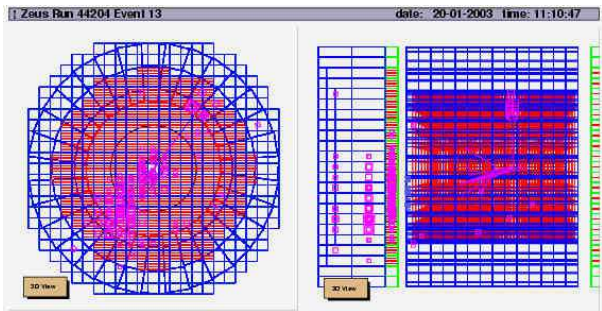}
\caption{Straight-forward projections of the 3-dimensional wireframe
  image into the $XY$ (left) and $ZX$ planes (right).}
\label{fig:straightProjection}
\end{figure}
\begin{figure}
\includegraphics[width=80mm]{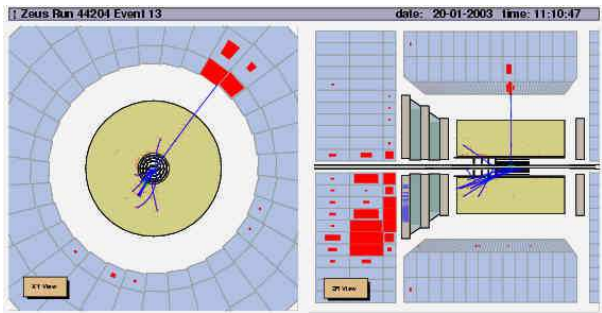}
\caption{Layered projections in $XY$ (left) and $ZR$ (right).}
\label{fig:layered}
\end{figure}
Standard projections in the straight-forward sense, e.g. along the
beam line or transverse to it, present detector and event in such a
way that coordinates could be measured, and every item is in principle
visible (fig.~\ref{fig:straightProjection}). However, straight-forward
projections can be very complex to look at in a multi-component
detector, and do not go all too well with the cylindrical geometry of
many colliding-beam experiments.

Very favourable for physics analysis are {\it layered projections},
which make use of filled areas to show geometry and event but
arrange them in suitable layers such that all relevant information is
visible (fig.~\ref{fig:layered}). An excellent review on how to design
useful projections for cylindrical detector designs has been given
in~\cite{drevermann}. The ideal system should integrate all three
viewing variants.

\section{Architecture And Design}
The ROOT system~\cite{root} was chosen as underlying graphics, GUI and
persistency package, since it provides drawing capabilities both for
2-dimensional graphics primitives as well as 3-dimensional graphics in
wireframe mode and with hidden lines and surface removal.  The ZeVis
package is structured such that the client application depends only on
one common library, which contains the geometry and event classes.
Other libraries contain the classes to build the geometry and event in
ROOT format from the ZEUS internal data model, and the server
application which will be discussed later.

The data structures are designed using the builtin ROOT containers, in
particular the {\it TClonesArray}. The geometry, which normally needs
to be loaded only once per session, is designed such that it can be
drawn quickly without further modification, hence the data structures
are all directly derived from ROOT geometry classes which are able to
draw themselves. The present size of the geometry file is about 450~kB.

For the event, compact size is important since a typical event display
session will involve access to many events. The data content is
therefore strongly reflecting the event information in the official
ZEUS mini-DST format, with about 30~kB/evt.

\begin{figure*}[t]
\centering
\includegraphics[width=135mm]{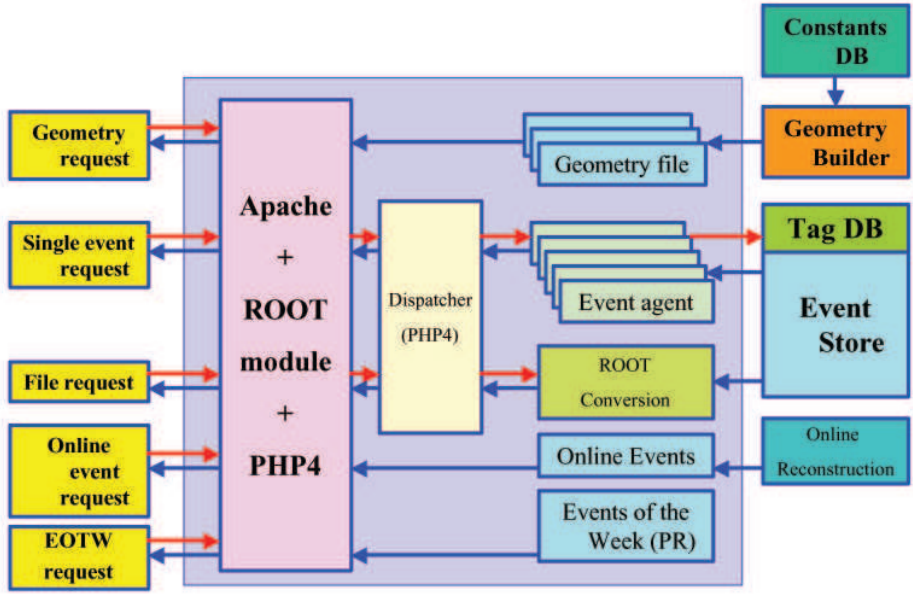}
\caption{Internal structure of the ZeVis server.}
\label{fig:serverInternal}
\end{figure*}
In layered projections, each geometry object acquires a 2-dimensional
shape which can be different in each projection, e.g. the drift
chamber outline is a circle in $XY$ view while it becomes two
rectangles in the $ZR$ view.\footnote{The $Z$ coordinate is defined by
  the proton beam direction, $X$ and $Y$ are transverse to it.
  $R=\sqrt{X^2+Y^2}$, signed according to azimuthal region.}  The
parallel implementation of 3-dimensional and 2-dimensional
representations is achieved by overloading the {\it Paint()} member
function in the derived geometry classes.

A fish-eye view introduces a nonlinear transformation of radius and
$Z$ coordinate in the layered $XY$ and $ZR$ projections, with the aim
to enable simultaneous inspection of inner (e.g. micro-vertex detector) and
outer tracking components (muon system) within the same picture. It is
implemented by overloading the {\it TView::WCtoNDC} member function of
ROOT.

\section{The Server}
The central server provides the access
to the geometry and event data for the client program which may be running
somewhere in the internet. It provides the following functionalities:
\begin{itemize}
\item detector geometry service
\item single event service
\item download of event files
\item access to online events
\item access to {\it events of the week} scan results (see explanation below)
\end{itemize}
\subsection{Internal Structure}
The internal structure of the ZeVis server is shown in
fig.~\ref{fig:serverInternal}. The server is based on a normal Apache
application~\cite{apache} which has been extended by a ROOT module and
PHP4. On the client side, requests to the server are handled by the
{\it TWebFile} class embedded in ROOT which has been slightly extended
to allow for passing of parameters and version qualifiers.  The server
operates its own cache, where a copy of each created event file is
kept for some time. A repeated request from a client will be satisfied
with the cached result of an identical earlier request if it still
exists.  Different components of the server provide the functionality
described above.

\subsection{Geometry Service}
The detector geometry is normally requested by the client
automatically on startup. The geometry resides statically on the
server, and is subject to version control. Each client binary
has the schema version it is based upon compiled in, and requests the
geometry explicitly in the required version. The server keeps the
geometry in several currently supported schema versions, which have
been generated offline.

\subsection{Single Event Service}
Particular care has been invested to provide a prompt return of single
events. The client requests a single event specified by run number and
event number, and a server application named {\it dispatcher} passes
it to a pre-initialized {\it event agent}, which uses the tag
database~\cite{zes1,zes2} to obtain the whereabouts of the event,
retrieves it from the event store and converts it into ROOT format
on-the-fly. The dispatcher then returns the event to the client. The
event agent is based on a normal ZEUS analysis application and has
access to all standard analysis code. It also uses the normal I/O
layers of all ZEUS applications, which initiate automatically staging
of a tape file if it should become necessary through the {\it
  dCache}~\cite{dcache}.  Since the initialization phase would
introduce a sizeable latency, event agents are always pre-initialized
and work in daemon fashion.  As an individual process could be kept
busy for some time if e.g. a file needs to be staged, several event
agents are operated in parallel. Similar as with the geometry, the server
supports several schema versions for event information. Typically, ten
event agents are working in parallel for each supported schema.
Besides run number and event number, the client can specify cuts on
physical quantities of the event, which the server evaluates as a
query on the tag database. Thus, the client is able to select e.g.
only DIS events in a certain kinematical range, or only diffractive
events with $J/\psi$ candidates in the muon channel, etc.

\subsection{Other Services}
In the download service, the client specifies the name of an event
file in the event store in one of several formats, which the server
converts into ROOT format on the fly. Also raw data files can be
requested, which the server passes through an on-demand
reconstruction.

During data taking, the ZEUS online system splits random events off
the stream with a typical frequency of 0.1~Hz. These events are passed
through an automatic reconstruction on a dedicated machine, converted
to ROOT and placed within the web area of the server. In online mode,
the client retrieves the latest online event periodically (this has
become a standard data quality control measure in the ZEUS control
room).

In regular running mode, teams of specialists scan preselected ZEUS
events visually and produce a selection named {\it events of the
  week}. The client application can retrieve this selection file
directly from the server, which is also a useful mode for public
relation purposes.

\subsection{Hardware and Performance}
The present event display server hardware is a dual processor PC with
Intel Xeon 2.2~GHz CPU and a fast IDE RAID controller with 240~GB of
disk space, connected by a gigabit ethernet interface.

\begin{figure}
\includegraphics[width=80mm]{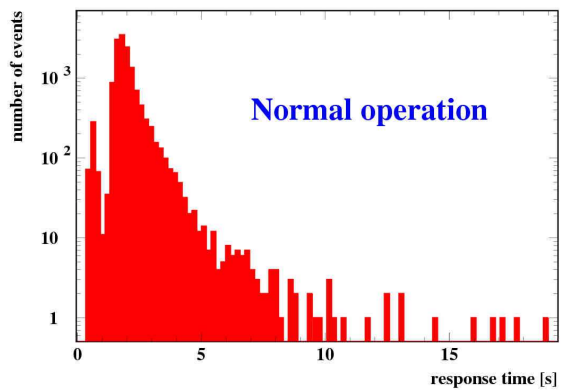}
\caption{Histogram of server response time under normal conditions.}
\label{fig:server1}
\end{figure}
\begin{figure}
\includegraphics[width=80mm]{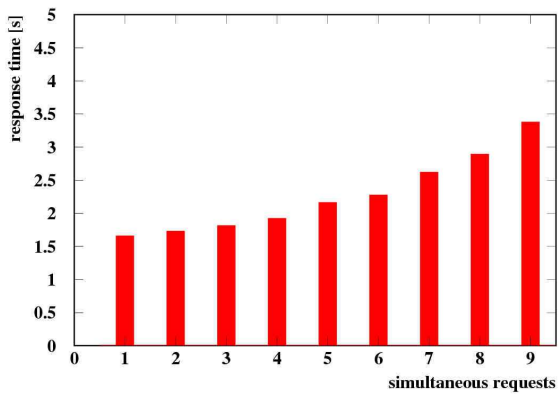}
\caption{Mean server response time as a function of request
  concurrency during a stress test.}
\label{fig:server2}
\end{figure}
The performance of the server under varying load conditions has been
investigated systematically. Figure~\ref{fig:server1} shows the
histogrammed response time of non-identical events requested in random
order, with the server cache deactivated. The response time is
measured on the server and includes access to the tag database and
event store, event analysis and conversion to ROOT format, but not the
network transfer to the client (which is almost negligible for 30~kB
event size on a normal local area network which can easily transfer
several MB/s).  The average response time is less than 2s, and will be
even shorter when the event happens to be in the cache. There are no
significant tails in the distribution. Figure~\ref{fig:server2} shows
the behaviour under an artificial stress test. While the response time
increases with increasing concurrency of demands, it is still below 4s
with nine simultaneous requests. With up to four requests in parallel,
hardly any degradation of performance is noticeable. This stress test
is however of rather artificial nature, since during normal operation,
concurrency of requests has hardly been observed at all.

\begin{figure*}
\includegraphics[width=150mm]{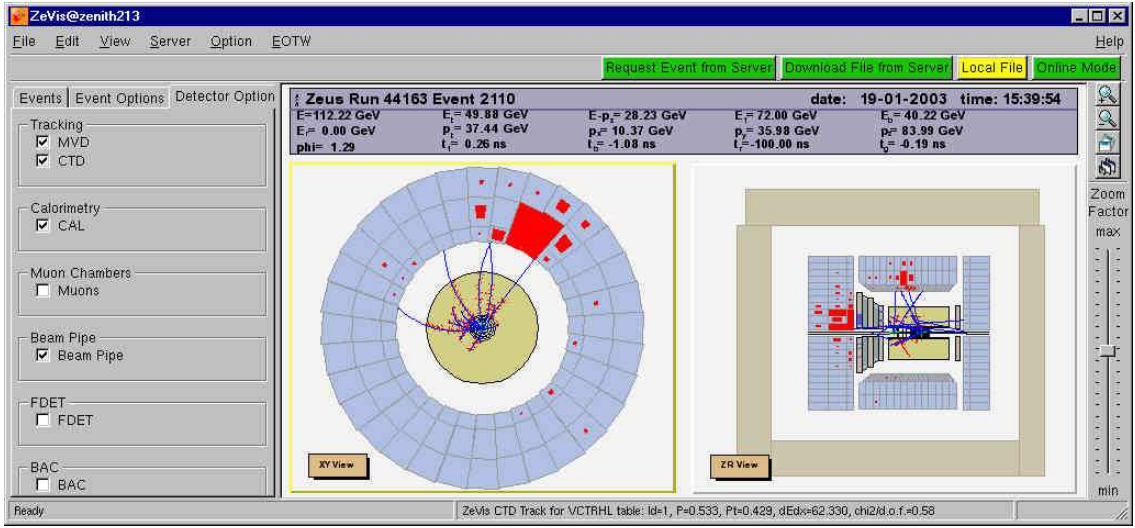}
\caption{Layout of the ZeVis GUI.}
\label{fig:gui}
\end{figure*}
\section{The Graphical User Interface (GUI)}
The general layout of the GUI is shown in fig.~\ref{fig:gui}. The
largest item is the event display canvas which consists of two pads
containing two selectable views of the event, and the header. Above
the canvas, four radio buttons allow selection of the event retrieval
mode, where the choice can be made between single event request, file
download, local and online modes. On the left side of the display
canvas, three register cards contain toggles for the display of
individual detector components and the corresponding event
information, and permit steering of the event retrieval, which will be
discussed in more detail below. The bottom edge of the frame shows the
general status on the left, and a description of the object currently
below the mouse pointer (in this case a track reconstructed in the
drift chamber). The right side of the display canvas features buttons
for zoom functions, and a slider to adjust the zoom factor. A set of
pull-down menus gives access to less frequently used options. The
mouse buttons can be used to pick displayed objects and inspect their
properties, in extension of a builtin ROOT feature. The context menu
also allows changing between the view modes.

\begin{figure}
\includegraphics[width=60mm]{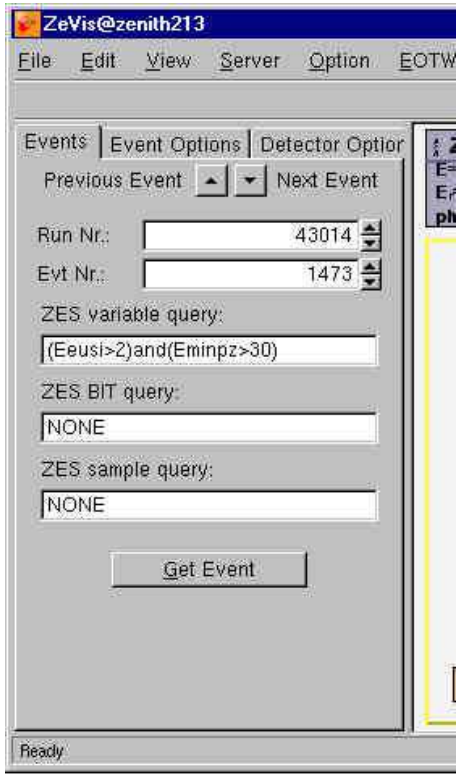}
\caption{GUI panel for single event requests.}
\label{fig:eventSelect}
\end{figure}
The steering panel for single event requests is shown in
fig.~\ref{fig:eventSelect}. The minimal specification consists of run
and event number.  Optionally, a query on event tag variables can be
entered in three fields, whereupon the server returns the first event
satisfying the query with an event number equal or greater than the
one specified. At the top, arrow-marked buttons allow navigation to
the next or previous event. In presence of a query, the server will
provide the next or previous event within the scope of the query.

\begin{figure}
\includegraphics[width=60mm]{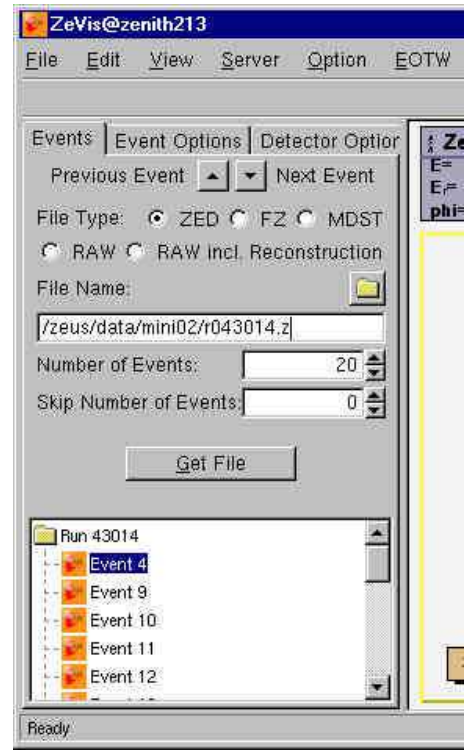}
\caption{GUI panel for the download of event files.}
\label{fig:download}
\end{figure}
A different panel (fig.~\ref{fig:download}) is used for requesting
complete files from the server in various formats. The file name can
be chosen with a file browser panel if the event store inventory is
visible from the host on which the client is
running. Figure~\ref{fig:download} also shows the graphical event
directory tree, which allows to move directly to any event in the
downloaded file.

\begin{figure*}
\includegraphics[width=150mm]{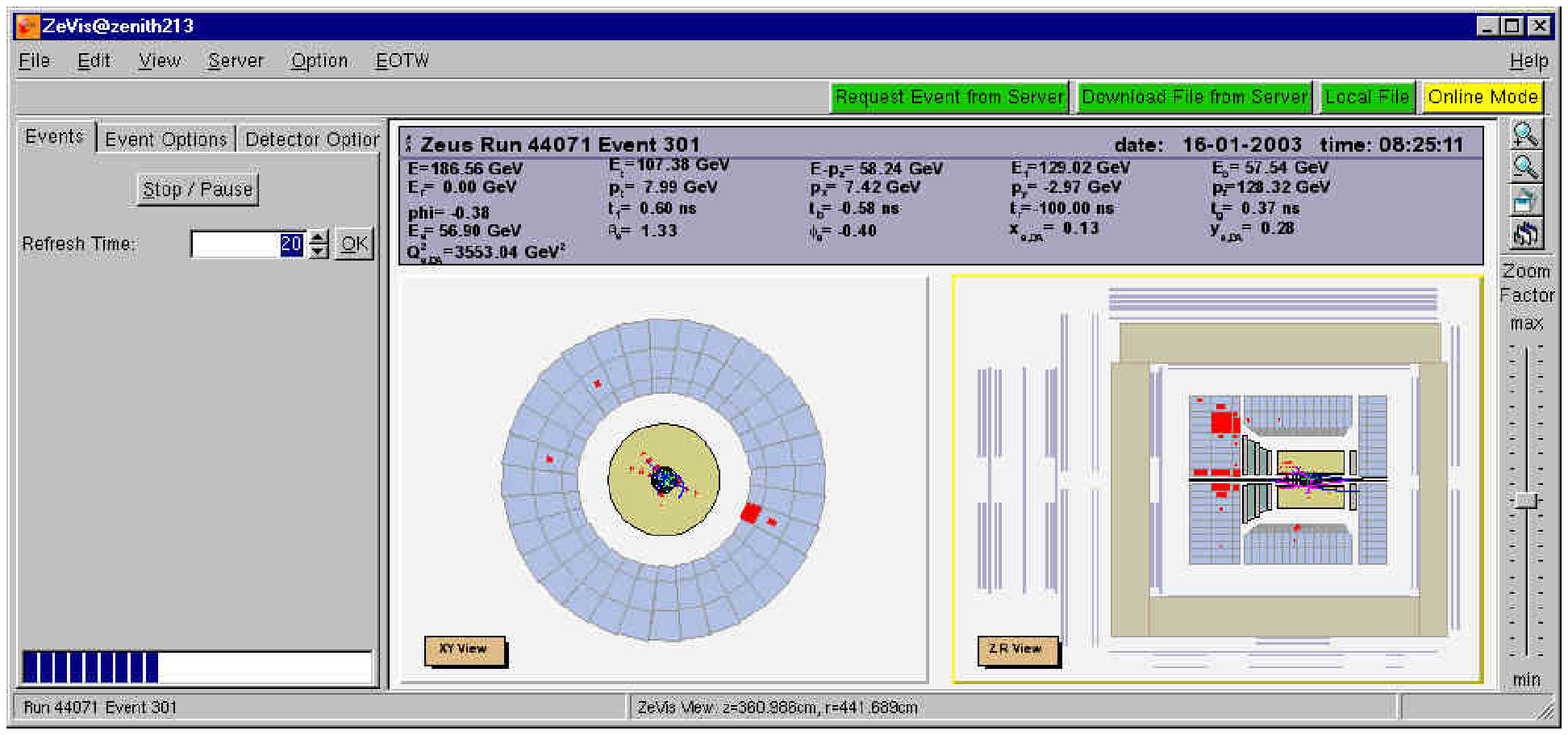}
\caption{GUI panel in the online mode.}
\label{fig:onlineEvt}
\end{figure*}
The GUI for the online mode is shown is fig.~\ref{fig:onlineEvt}. In this mode,
the client periodically requests the latest online event from the
server (here every 20 seconds).

\begin{figure*}
\includegraphics[width=150mm]{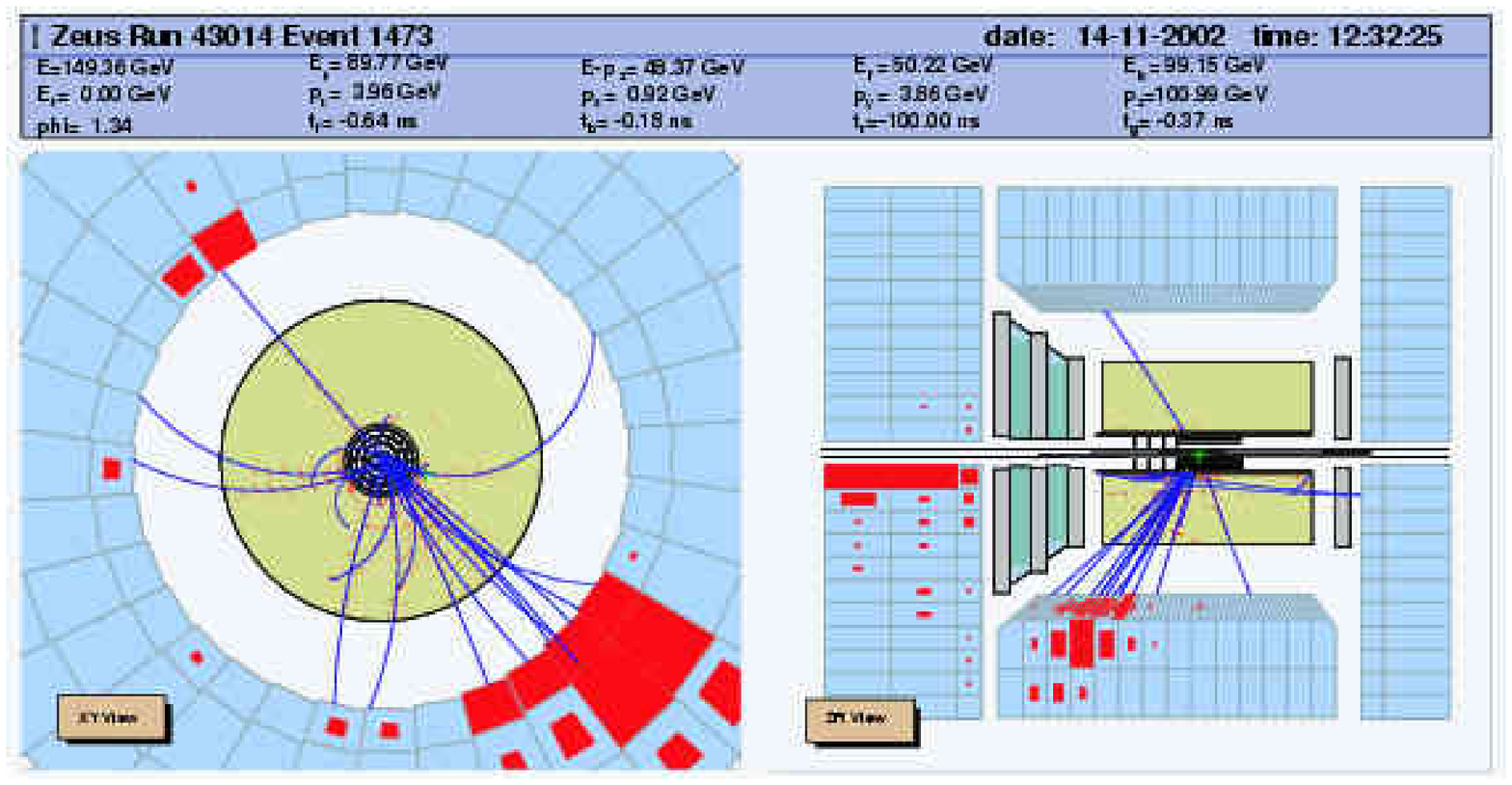}
\caption{Event display with layered projections in $XY$ and $ZR$.}
\label{fig:xyAndRz}
\end{figure*}
\section{Visualization of Events}
A very typical event display for daily work is shown in
fig.~\ref{fig:xyAndRz}. It shows layered projections in $XY$ (left) and
$ZR$ (right). The calorimeter energies are represented by the size of
the shapes within the cell images, and are summed over all
corresponding cells in $Z$ in case of the $XY$ view, and azimuth
$\phi$ in case of the $ZR$ view.

\begin{figure*}
\includegraphics[width=150mm]{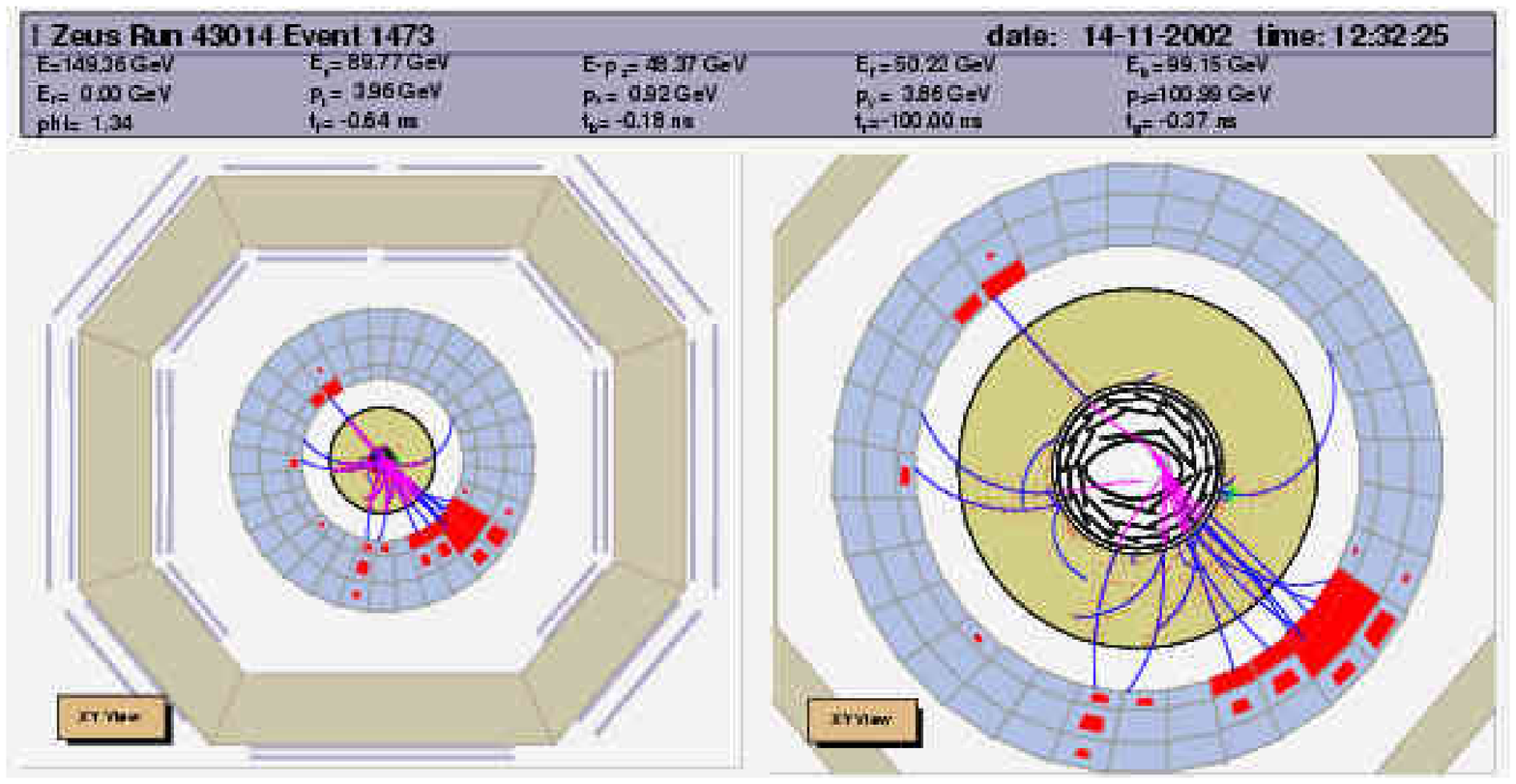}
\caption{Linear $XY$ view (left) in comparison with fish-eye view (right).}
\label{fig:fisheye}
\end{figure*}
The fish-eye view allows simultaneous inspection of the inner MVD
tracking layers together with the outer components
(fig.~\ref{fig:fisheye}). The fish-eye transformation blows up the
inner part of the detector, it is available in both $XY$ and $ZR$
projections.

\begin{figure*}
\includegraphics[width=150mm]{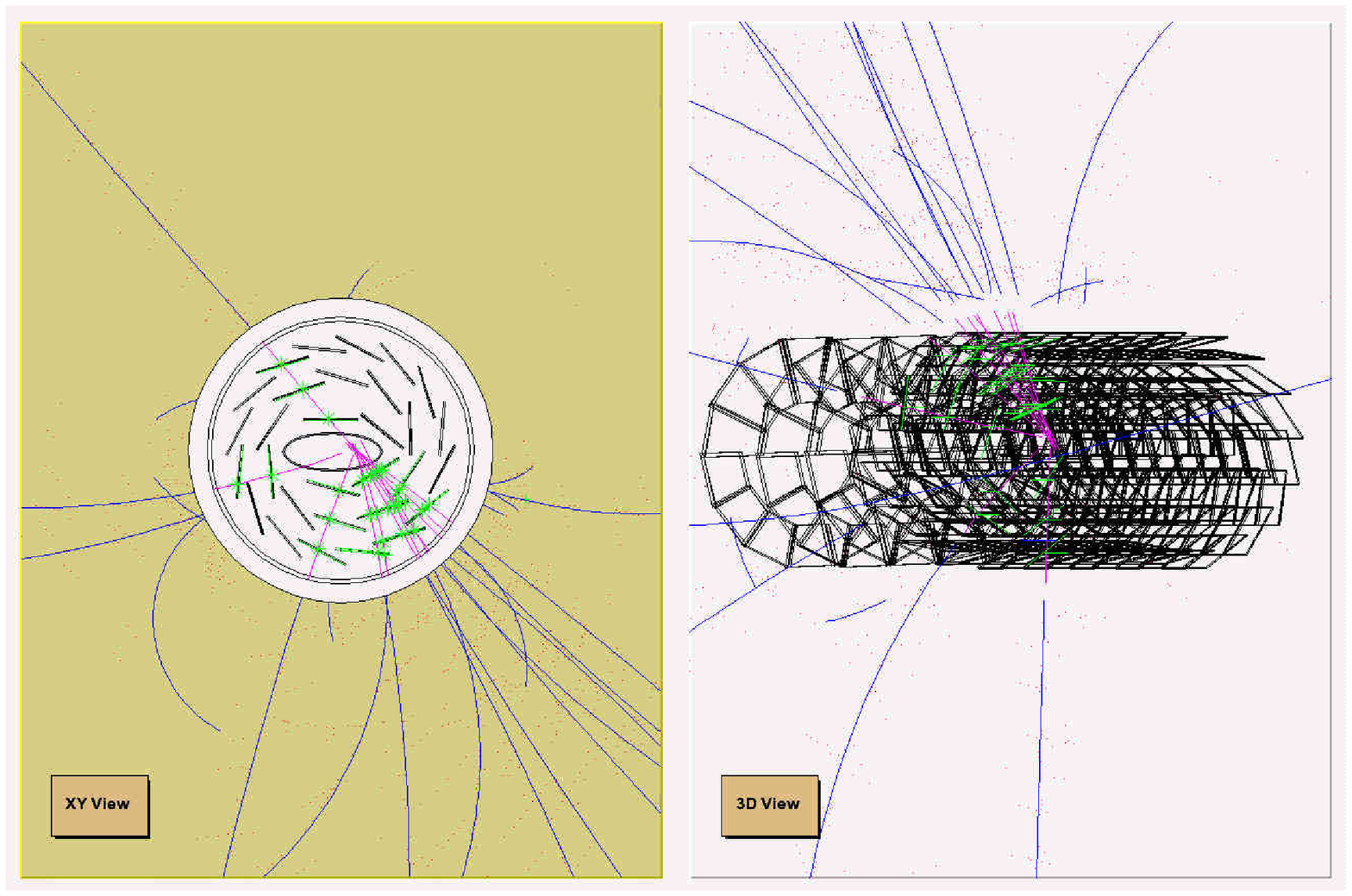}
\caption{$XY$ view (left) and 3-dimensional view (right) of the MVD.}
\label{fig:mvdHits}
\end{figure*}
The micro-vertex detector benefits particularly from the capability to
display 2-dimensional and 3-dimensional views together.
Figure~\ref{fig:mvdHits} shows the MVD simultaneously in $XY$
projection and at an arbitrary spatial viewing angle.

\begin{figure*}
\includegraphics[width=150mm]{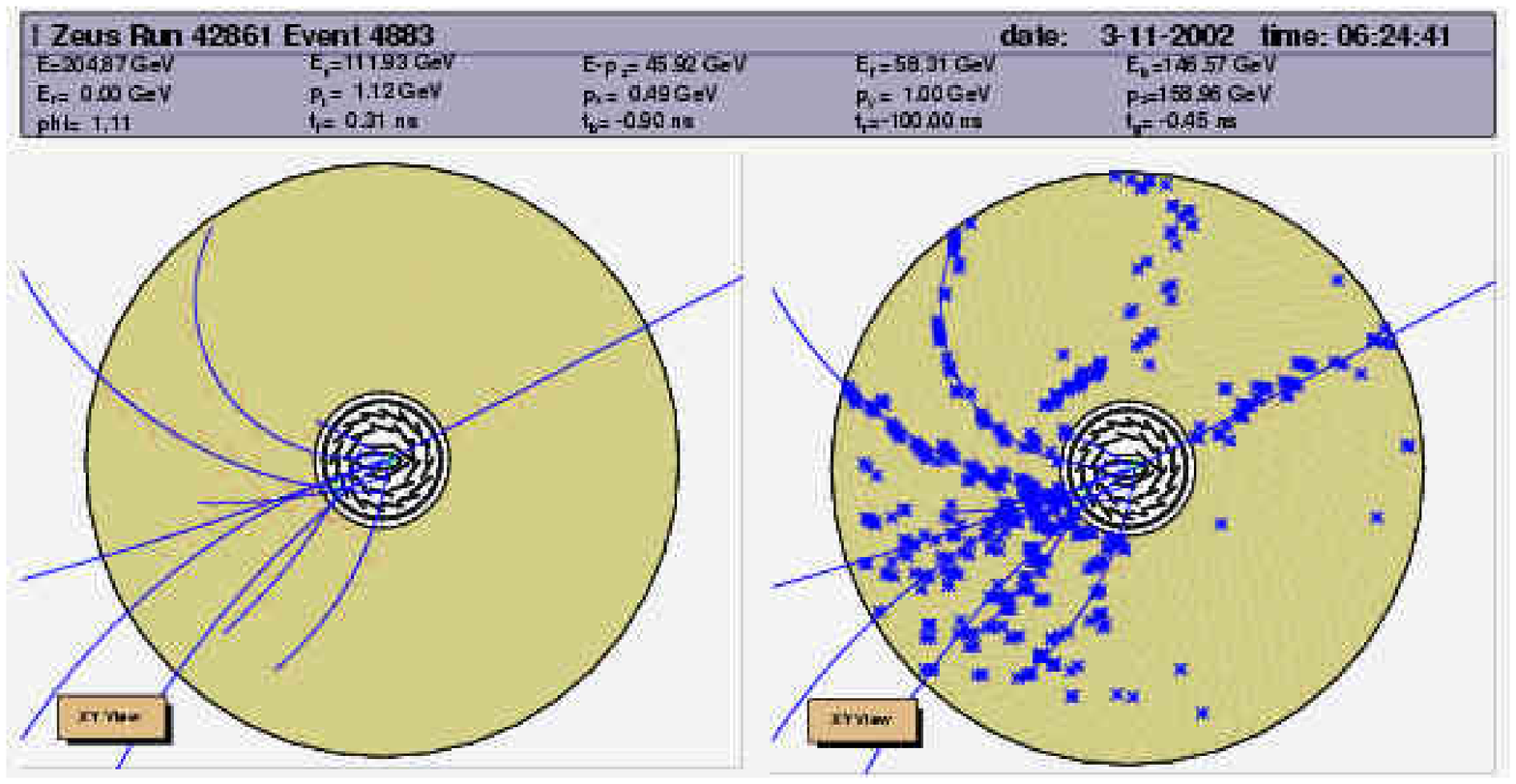}
\caption{Drift chamber event in normal mode (left) and with sense
  wires and raw hits activated (right).}
\label{fig:rawHits}
\end{figure*}
\begin{figure*}
\includegraphics[width=150mm]{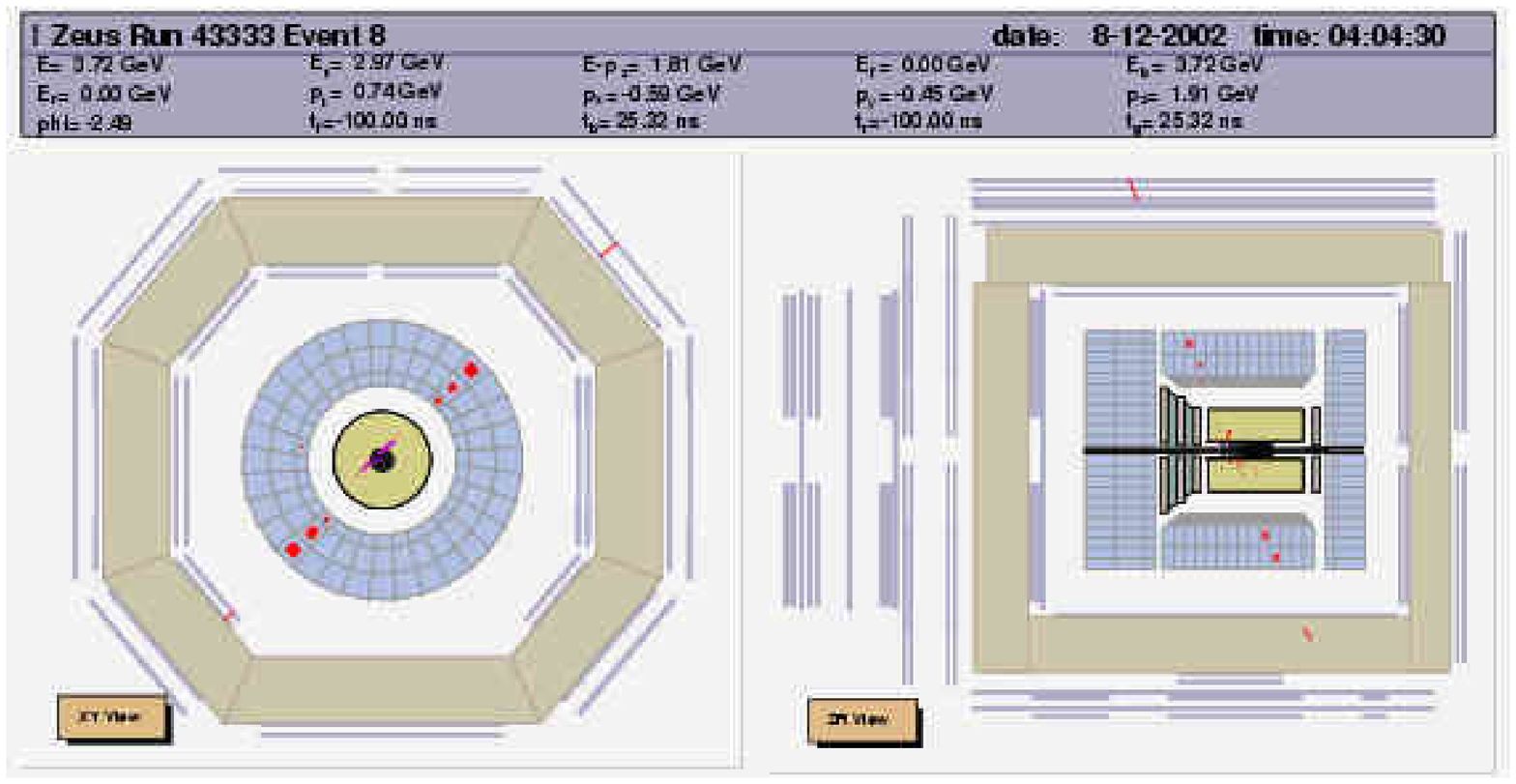}
\caption{Display of a cosmic muon event, with track segments in the
  muon chambers.}
\label{fig:muonHits}
\end{figure*}
Special modes enable display of the raw hit information, as illustrated
in fig.~\ref{fig:rawHits} in case of the central drift chamber. A
display of a cosmic muon event with track segments in the muon
chambers is shown in fig.~\ref{fig:muonHits}.

Three-dimensional views with hidden lines and surface removal are
possible through the special viewers {\it X3D} and {\it OpenGL} which
are integrated in view. Figure~\ref{fig:detector-3D} has been obtained
using OpenGL. With these modes, live rotations are possible, and a
suitable hardware acceleration, as available in the graphics adapters
of most contemporary PCs, provides instantaneous response.

\section{Summary}
We have developed and deployed a client-server event display for the
ZEUS experiment, which is based on the ROOT framework. Its main
features are a portable, lightweight client, smooth integration of
2-dimensional and 3-dimensional views and an ergonomic graphical user
interface. The performance of the server is excellent and leaves
plenty of reserves for future increase in usage.

\vspace{8mm}
\begin{acknowledgments}
  It is a pleasure to thank the ZEUS detector and physics groups for
  their help and many useful suggestions about the most efficent
  implementation of detector geometry and event data, and the ROOT
  team for help and advice in connection with ROOT, as well as for
  providing the system in the first place. We would also like to thank
  our summer students Aleksandra Adametz, Valentin Sipica and Ildar
  Tamendarov for their contributions to the program.
\end{acknowledgments}


\begin{thebibliography}{9}   

\bibitem{drevermann} H.~Drevermann, B.S. Nilsson and D. Kuhn,
 {\it Is there a Future for Event Display?}, Proceedings of Cern
 Computing School, L'Aquila 1992, 102-134.

\bibitem{root} http://root.cern.ch

\bibitem{apache} http://www.apache.org

\bibitem{zes1}
L.~Bauerdick et al., {\it Event Indexing Systems For Efficient Selection
and Analysis of HERA Data}, Comput.Phys.Commun.137:236-246,2001.

\bibitem{zes2}
U.~Fricke, {\it Upgrade of the ZEUS OO Tag Database for Physics Analysis at
HERA}, Proceedings of the CHEP01 Conference, Beijing 2001, 252-255.

\bibitem{dcache} http://www.dcache.org


\end{thebibliography}

\end{document}